\documentclass[twocolumn,epjc3]{svjour3}
\smartqed  % flush right qed marks, e.g. at end of proof
\RequirePackage{graphicx,amssymb,color}
\journalname{Eur. Phys. J. C}
\newcommand{\black}{\color{black}}
\newcommand{\red}{\black}
\newcommand{\green}{\black}

\begin{document}

\title{The local-filament pattern in the anomalous transparency of the
Universe for energetic gamma rays\thanksref{t1}
}


\author{Sergey Troitsky\thanksref{e1,addr1}
}

\thankstext{t1}{This work was supported by the Russian Science
Foundation, grant 18-12-00258.}
%Grants or other notes
\thankstext{e1}{e-mail: st@ms2.inr.ac.ru}


\institute{Institute for Nuclear Research of the Russian Academy of
Sciences, 60th October Anniversary Prospect 7a, 117312, Moscow,
Russia \label{addr1}
%           Second address \label{addr2}
%           \emph{Present Address:} if needed\label{addr3}
}

\date{Received: date / Accepted: date}

\maketitle

\begin{abstract}
The propagation length of high-energy photons through the Universe is
limited by $e^{+}e^{-}$ pair production on the
extragalactic background radiation. Previous studies
reported discrepancies between predicted and observed
attenuation, suggesting
explanations in terms of new physics. However,
these effects are dominated by a limited number of observed sources, while
many do not show any discrepancy. Here, we
consider the distribution in the sky of these apparently anomalous
objects, selected in two very different approaches: the study of
unphysical hardenings at distance-dependent energies in deabsorbed spectra
of TeV blazars, and the observation of ultra-high-energy air
showers from the directions of BL Lac type objects.
In both cases, directions to the anomalous sources follow the projected
local distribution of galaxies: all the distant
sources, contributing to the anomalies, are seen through the local
filament. This \black matches \black the prediction of the proposed earlier
explanation of the anomalies based on mixing of photons with axion-like
particles in the filament's magnetic field.
\red
For ultra-high energies, this axion interpretation may be tested
by the search of primary gamma rays.\black
% \keywords{First keyword \and Second keyword \and More}
\end{abstract}

\section{Introduction}
\label{sec:intro}
Distant active
galaxies, blazars, are among the strongest sources in the very-high-energy
(VHE; above $\sim 100$~GeV) gamma-ray band. They were detected even with
the first modest instruments, despite the predicted \cite{Nikishov}
attenuation of the flux: energetic gamma rays
from distant sources interact with soft photons constituting the
extragalactic background light (EBL) and produce $e^{+}e^{-}$ pairs. Though
direct observations of EBL suffer from uncertainties, lower limits on
the EBL are firm  \cite{EBL-review,EBL-recent-review} and allow
one to take the absorption into account.
Gamma-ray spectra of some distant blazars, \textit{deabsorbed} with these
models, were found to be too hard compared to those of similar nearby
objects. This problem was called ``the
infrared/TeV crisis''  \cite{crisis} in 2000.
\red Several solutions have been proposed, see below,
though there is still no generally
accepted solution to the issue. \black Modern imaging atmospheric
Cherenkov telescopes (IACTs) enlarged the amount of blazars detected in
VHE to several dozens, opening the way to analyzing statistical samples.
The anomalous behavior of a VHE blazar has been identified
\cite{HornsMeyer,gamma} as hardening, that is an upward change of the
spectral slope, in the spectrum deabsorbed with the ultimate lower-limit
EBL. For blazars located at different distances from the observer $D$, the
absorption on EBL becomes important at different energies $E_{0}(D)$, and
the hardenings are observed precisely at these energies $E_{0}$,
indicating an incorrect account of the absorption. This was termed
``anomalous transparency of the Universe''. However, most recent studies,
based on larger samples of blazars and better measured distances to them,
indicate that the overall significance of the anomalous transparency seen
in VHE blazars is modest
\cite{gamma2,fermi2,veritas}\footnote{References~\cite{Biteau,magic,hess}
reached similar conclusions from analyses treating spectra of one and the
same object observed multiple times as statistically independent data, see
e.g.\ Refs.~\cite{gamma2,ST-PRD} for discussions.}. Spectra of many newly
discovered blazars are consistent with the pair-production attenuation for
low-EBL models, but several bright objects observed previously continue to
demonstrate the anomaly. In addition, new sources with anomalous
hardenings have been found, see Ref.~\cite{gamma2} for the list and
references.

On the same timescale of decades, another puzzling effect has been
observed and widely discussed, the directional correlation of some of
ultra-high-energy (UHE; above $\sim 10^{18}$~eV) cosmic-ray air showers
with BL~Lac type objects, a subclass of blazars. The correlation was
found \cite{BLL} for the published set of cosmic-ray events with the
primary energy $E>10^{19}$~eV detected by the High Resolution Fly's Eye
(HiRes) fluorescence air-shower detector in the stereoscopic
mode \cite{HiRes271}. Given the deflection of charged cosmic-ray particles
in astrophysical magnetic fields, the observed directional coincidence
implies neutral primaries. However, assuming standard physics, no neutral
particle with that high energy can reach us from the distances at which BL
Lacs are located \cite{TT-no-neutral} \red (except for neutrinos; see
e.g.\ Ref.~\cite{BLL-UHE-nu} for the discussion of a possible
BL~Lac -- UHE cosmic rays -- neutrino connection)\black. Subsequently, the
HiRes collaboration has confirmed \cite{BLL-HiRes} the observation and has
extended it to lower energies, $E \gtrsim 10^{18}$~eV \red (see also
Ref.~\cite{FarrarBL} for an independent analysis of the same
$E>10^{19}$~eV data by a different method, confirming the effect with a
stronger estimated significance)\black. The angular resolution of the
HiRes experiment in the stereo mode, $0.6^{\circ}$, remains unsurpassed,
and the correlation has not been tested with data of modern fluorescence
air-shower detectors yet.

Theoretical approaches to the explanation of the anomalous transparency
exploit either the suppression of the pair production or the assumption
that the observed photons do not come from the source but are born much
closer to the observer. The first option is possible only in models with
the Lorentz-invariance violation\red, see Ref.~\cite{LIV-rev} for a recent
review. However, for the required parameters, the Lorentz-invariance
violation \black would suppress also the development of air showers in the
atmosphere and therefore make the VHE photons invisible for Cherenkov
telescopes \cite{no-LIV1,no-LIV2}, contrary to observations (see however
Ref.~\cite{Hassan} \red for a model overcoming the limit\black). The second
option may be realized in two approaches, ``cascade'' and
``conservation''. Cascades can develop in the cosmic background fields and
radiation, starting either from a VHE photon \cite{Dzhatdoev} or from an
accompanying UHE proton emitted by the same source \cite{Kusenko} \red
(see also Refs.\
\cite{Essey:2010er,Murase:2011cy,Takami:2013gfa,Tavecchio:2013fwa,%
Zheng:2015gfw,Tavecchio:2018nrt})\black; what we detect at the Earth may
be secondary particles produced nearby. Conservation means that the photon
converts near the source to some particle, which does not produce pairs on
EBL, then this particle travels unattenuated and reconverts back to a
photon close to us. A viable mechanism involves oscillations
\cite{RaffeltStodolsky} of photons to hypothetical axion-like particles
(ALPs) in external magnetic fields \cite{DARMA,Serpico,Csaba,ST-mini}.
Amazingly, the very same ALP-based mechanism can \cite{FRT} explain in a
consistent way the correlations of UHE showers with BL Lacs, which
otherwise remain misterious.

The
ALP-photon
mixing requires magnetic fields, while the cascades get distorted by the
fields so that secondary charged particles, and hence photons born in
their interactions, no longer point to the original source of the
high-energy emission. Therefore, the two scenarios predict opposite
patterns of anisotropy in the distribution of the effect strength over the
sky: in the cascade case, anomalies are expected to follow low-field
regions, while in the ALP case, regions of higher field are favoured.
It has been pointed out that the VHE anomalies may be
explained by the ALP-gamma conversion in the Galactic magnetic field and
the corresponding anisotropy might be related to the
Galactic
plane \cite{Serpico}. For UHE photons, the conversion in the Galaxy is
suppressed, but local extragalactic structures, small-scale filaments,
provide for the required conditions; conversion of VHE photons in the
filament field is also possible for certain ALP parameters \cite{FRT}.

Several attempts to find deviations from isotropy in the anomalous
transparency have been made.
\black
Ref.~\cite{Furniss}, based on the Fermi-LAT data, studied the
correlation between VHE candidate sources and the line-of-sight
``voidedness'' parameter, reporting a weak indication that sources with
hard spectra are seen through voids in the large-scale structure. Their
analysis did not select the sources with anomalous spectra and probed the
matter distribution in the distant Universe at $z>0.05$,
complementary to the Local supercluster we study here.

\black
In Ref.~\cite{Serpico}, the distribution in
the sky of a few distant VHE blazars, known at that time, was
compared to the maps of the photon-ALP conversion probability calculated
with three models of the Galactic magnetic field. Visually, a possible
correlation with the high-probability regions was pointed out for one
\cite{HMR} of the field models. Subsequently, in Ref.~\cite{ST-PRD}, a
similar study has been repeated with the same field model and an enlarged
sample of VHE blazars, qualitatively confirming the trend. However, the
model \cite{HMR} of the field is outdated and the test is not perfect
because it does not distinguish truly anomalous objects. More recent
magnetic-field models predict \cite{Horns1207.0776,WBrun} different
patterns of the conversion probability in the Milky Way, so the anisotropy
seen in Refs.~\cite{ST-PRD,Serpico}, if real, is not related to the
Galaxy. In Ref.~\cite{FRT}, it has been demonstrated that the cosmic rays
correlating to BL Lac type objects do not follow the HiRes exposure, but
no particular pattern was tested.

In the present work, we revisit the
question of anisotropy in the anomalous transparency for both VHE and UHE
cases and concentrate on the pattern associated with the local filament.
\section{Analysis and results}
\label{sec:anal}
\subsection{Method}
\label{sec:anal:method}
We use the weighted density of galaxies along the line of sight as a
tracer of the local large-scale structure. The method was used extensively
in the studies of cosmic-ray anisotropies at UHE, see e.g.\
Refs.~\cite{UHE-LSS1,UHE-LSS2,UHE-LSS3}. The starting point is a
flux-limited catalog of galaxies for which we use the 2MASS Redshift
Survey (2MRS), Refs.~\cite{2MASS1,2MASS2}.
Weights~\cite{sliding-old,Hylke} are chosen in such a way that
the incompleteness of the catalog is compensated.

\black The 2MRS catalog traces the large-scale structure up to distances of
250~Mpc. Photons of different energies have different mean free paths with
respect to the pair production, varying from a few Mpc for
$10^{19}$~eV UHE gamma rays to a few Gpc for the most distant VHE sources
observed at $\sim 100$~GeV. In the frameworks of the ALP scenario, nearby
regions with magnetic field serve as secondary sources for reconverted
photons. The flux from these secondary sources should be therefore
suppressed by the inverse distance squared, so we introduce \black
the $1/D_{L}^{2}$ weight, where $D_{L}$ is the luminosity distance to a
particular galaxy. \black This \black means that nearby galaxies contribute
more, so the weighted density is higher for the directions parallel to the
local filament in which the Milky Way galaxy resides, a $\sim$Mpc thick
sausage extending from the Virgo cluster to the Fornax cluster, see e.g.
Ref.~\cite{local-filament}.

The weighted density $f(l,b)$ is determined as
a function of the direction in the sky given by the Galactic coordinates
$(l,b)$ as described in Appendix~\ref{app:magic-numbers}; larger values of
$f$ therefore trace the local large-scale structure. We stress that the
local filament extends to $\sim 20$~Mpc from us while nearest blazars are
$\sim 150$~Mpc away, hence their distribution is completely independent of
the local structure traced by $f$. Following Ref.~\cite{UHE-LSS1}, we
calculate two sets of values, $\mathcal{S}=\{f(l_{i},b_{i})\}_{\rm S}$ for
the set $S$ of directions which are associated with the anomalous
transparency effects, and $\mathcal{B}=\{f(l_{i},b_{i})\}_{\rm B}$ for the
control sample $B$ of directions in the sky. The two sets of numbers,
$\mathcal{ S}$ and $\mathcal{ B}$, are then compared by means of the
Kolmogorov-Smirnov test which gives the probability that they are derived
from one and the same distribution.

This method requires to fix the angular smoothing scale when the function
$f(l,b)$ is calculated, see Appendix~\ref{app:magic-numbers}.
\green While in
cosmic-ray studies the scale was determined by expected deflections of
charged particles in cosmic magnetic fields, here we do not have such
a guidance.
While we are interested
in the interpretation involving the filament's magnetic fieds, they are
poorly measured and we do not know to which extent the number density of
galaxies traces the field. Hence we choose to treat the smoothing as a
free parameter of the model, see \ref{app:magic-numbers}, and
account for this freedom in a standard statistical
approach~\cite{TT-penalty} described in \ref{app:trials}.
\black

\subsection{Results for TeV hardenings}
\label{sec:anal:TeV}
\begin{figure}
\begin{center}
\includegraphics[width=0.9\columnwidth]{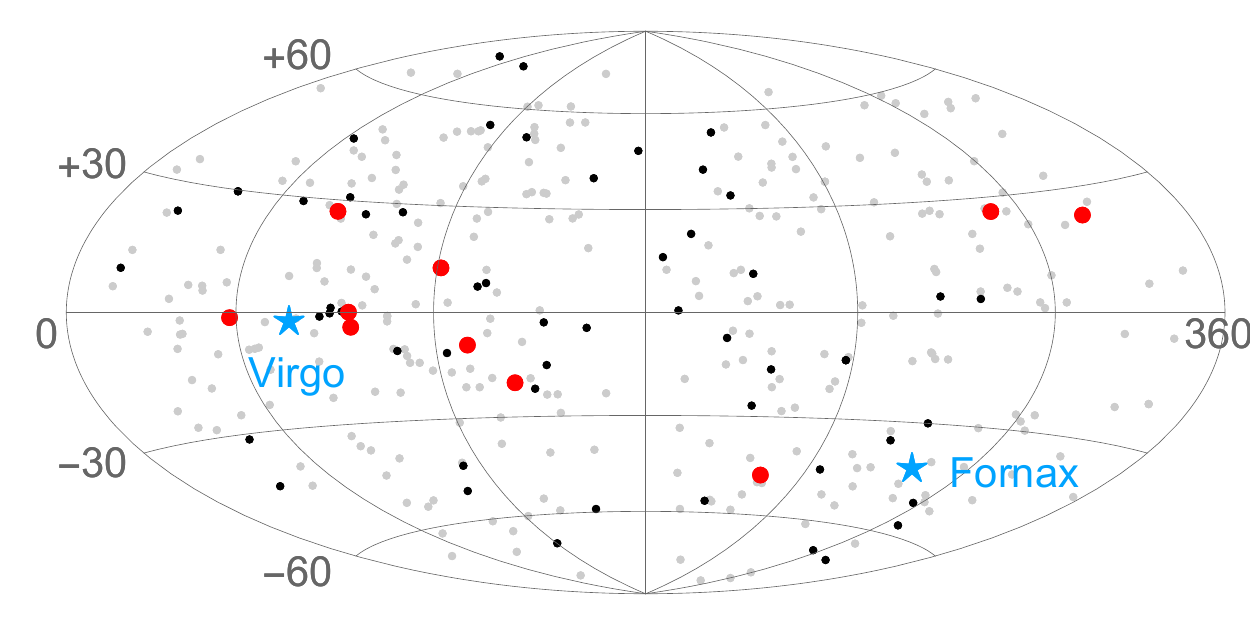}
\caption{
\label{fig:mapPBR-dots}
Objects with anomalous hardenings in VHE (red circles) together with other
objects from the TeVCat (black dots) and 3FHL (gray dots) samples in the
celestial sphere (supergalactic coordinates).
}
\end{center}
\end{figure}

For the VHE blazars, we start with the data sets of Ref.~\cite{gamma2}.
There, a sample of blazars with known redshifts $z$ was constructed
starting from the TeVCat catalog  \cite{TeVCat} for the sources observed
by
Imaging Atmospheric Cherenkov Telescopes (IACTs; 66 sources) and from the
3FHL catalog  \cite{3FHL} for those observed with Fermi LAT (307 sources
at
$z>0.2$)\footnote{Strict quality cuts applied in Ref.~\cite{gamma2}
removed several sources claimed as anomalous in previous studies.}. Only 26
and 5 of them, respectively, were detected at considerable opacities, and
8 and 2 have hardenings (upward spectral breaks at the energy
$E_{0}(D)$ inconsistent with zero at 68\% CL)
in the spectra deabsorbed with the most recent EBL
model~\cite{KorochkinEBLmodel}. These latter 10 objects, see
Table~\ref{tab:pbr} in Appendix~\ref{app:catalogs}, are considered here as
the signal sample $S_{V}$ of objects demonstrating anomalies, while the
control sample ${B}_{V}$ is constructed of 1000 randomly selected sets of
8 of 66 IACT and 2 of 307 Fermi-LAT sources.
The distribution of objects from the VHE samples in the sky are shown in
Fig.~\ref{fig:mapPBR-dots}.
The pre-trial Kolmogorov-Smirnov probability that the control and signal
samples are derived from one and the same distribution is 0.013 and
corresponds to the disk smoothing of $4^\circ$ (see
Appendix~\ref{app:magic-numbers} for details and Fig.~\ref{fig:trials} in
Appendix~\ref{app:trials} for the dependence of the local p-value on
the smoothing). The post-trial probability is 0.028.

\subsection{Results for EeV correlations}
\label{sec:anal:EeV}
\begin{figure}
\begin{center}
\includegraphics[width=0.9\columnwidth]{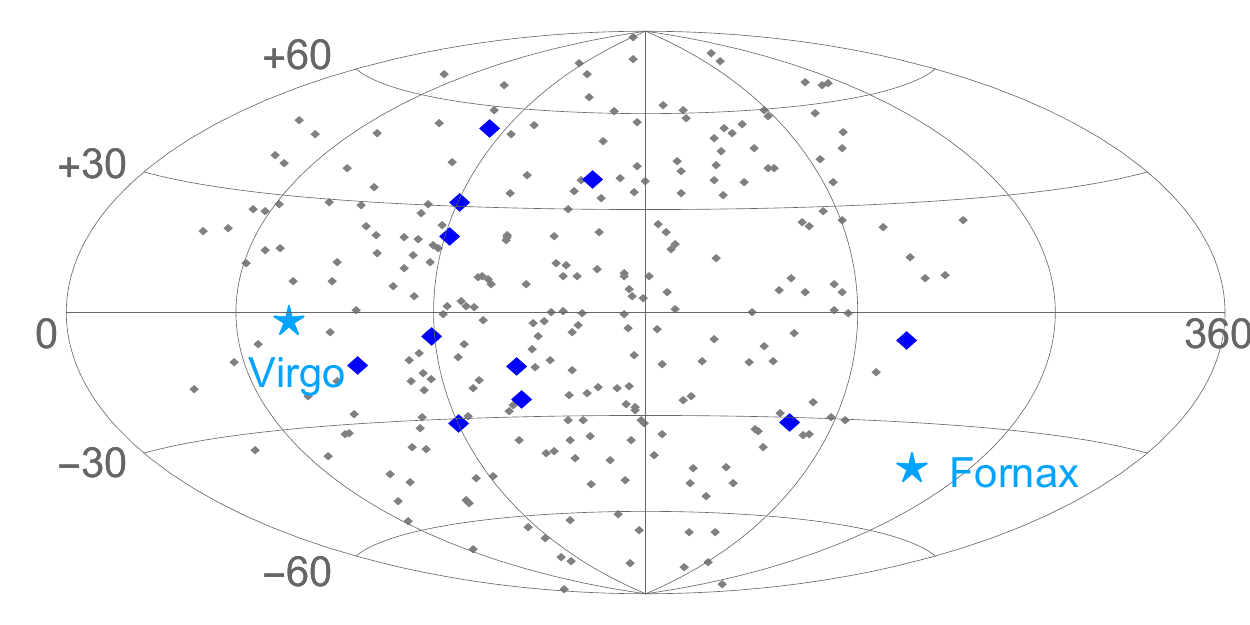}
\caption{
\label{fig:mapBL-dotsHS}
Arrival directions of HiRes stereo UHE air showers correlated with BL~Lac
type objects (blue diamonds) together with directions of other showers in
the sample (gray dots) in the celestial sphere (supergalactic coordinates).
}
\end{center}
\end{figure}

We now turn to the UHE case. Here, we start with Ref.~\cite{BLL} where a
certain number of HiRes air showers with reconstructed primary energies
above $10^{19}$~eV were found to correlate with BL~Lac type objects. Of
156 sources selected in previous studies, 11 were found to be within
$0.8^{\circ}$ from the shower arrival directions, while only 3 were
expected assuming isotropy. This angle was determined from the angular
resolution of the experiment, $0.6^{\circ}$, by a Monte-Carlo simulation
as maximizing the signal-to-noize ratio, assuming neutral primary
particles from BL Lac's. We consider these 11 directions as the signal
sample ${S}_{U}$, see Table~\ref{tab:bll} from
Appendix~\ref{app:catalogs}, for the list. The control sample
${B}_{U}$ is given by arrival directions of all 271 HiRes air
showers in the sample
\cite{HiRes271}, see Fig.~\ref{fig:mapBL-dotsHS}.
The pre-trial Kolmogorov-Smirnov p-value for these two samples is
$5.2 \times 10^{-4}$, achieved for the Gaussian smoothing at the
$2^{\circ}$ scale. The post-trial probability is $1.1 \times 10^{-3}$.

\begin{figure}
\begin{center}
\includegraphics[width=0.9\columnwidth]{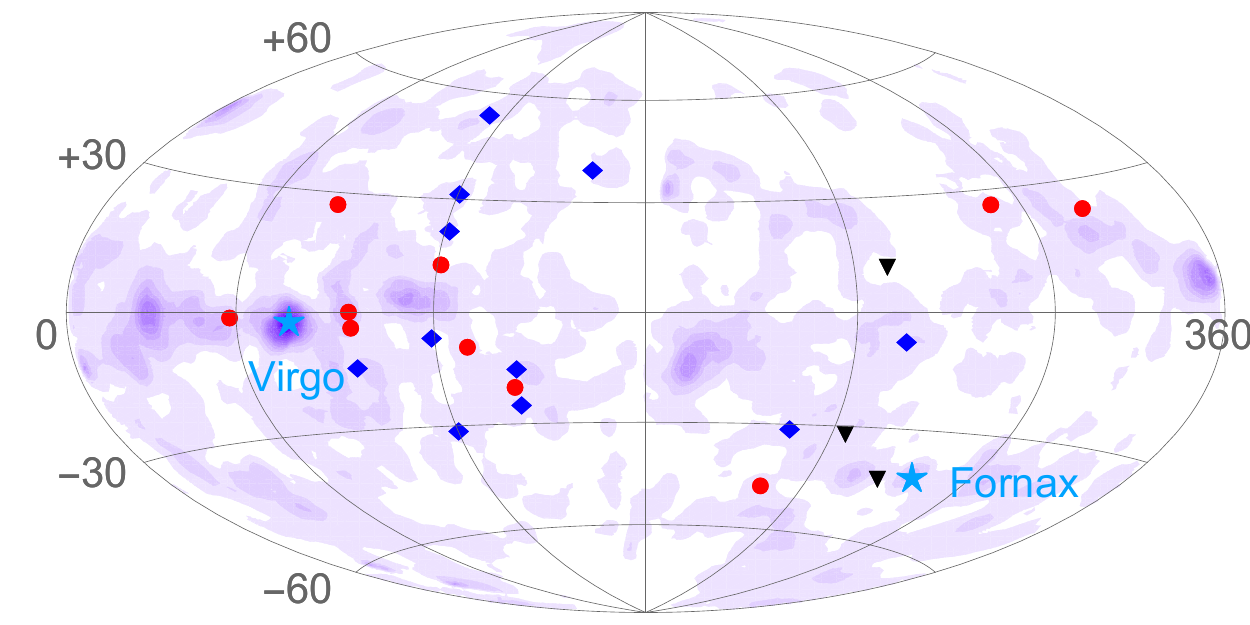}
\caption{
\label{fig:mapBLandPBRandLSS}
Objects with anomalous hardenings in VHE (red circles), HiRes stereo UHE
air showers correlated with BL~Lac type objects (blue diamonds) and
gamma-ray bursts detected in VHE (black triangles) together with the
weighted galaxy distribution (density plot) in the celestial sphere
(supergalactic coordinates). }
\end{center}
\end{figure}

Figure~\ref{fig:mapBLandPBRandLSS}
shows the distribution of anomalous directions in the sky superimposed
on the density plot of the projected weighted galaxy distribution.
In
both VHE and UHE cases, the signal sample follows the regions of higher
weighted densities. We address the difference in statistical significance
between the two samples below. The two
results represent independent tests of the anomalous
transparency at different energies. We
compare now combined signal and background samples, so that
the signal
distribution ${S}_{V+U}$ consists of directions of all 21
anomalous sources while the control sample ${B}_{V+U}$ is obtained
by taking 1000 random subsamples, each consisting of 8 of 66 TeVCat
blazars, 2 of 307 3FHL blazars and 11 of 271 HiRes arrival directions.
This results in the pre-trial $p_{1}=5.2 \times 10^{-5}$, again achieved
for the Gaussian smoothing at the $2^{\circ}$ scale. The post-trial
probability is $p=7.5 \times 10^{-5}$. Were the statistics Gaussian, this
would correspond to the \textbf{significance of 4.0 standard deviations
(post trial)}.

\section{Discussion and conclusions}
\label{sec:concl}
We see that, for both samples, the sources whose observation gave the
base for claims of the anomalous transparency are seen along the local
filament, through the regions of higher matter density, and probably higher
magnetic fields.
\black We tested explicitly that our results do not depend on the
maximal distance cut in the catalog, so they are indeed dominated by the
Local supercluster. Including only galaxies within the spheres of 20, 30,
50, 75, 100, 150 and 250~Mpc and repeating the statistical procedure, we
obtained roughly the same statistical significance. While the resulting
pre-trial p-value develops a shallow minimum around 30~Mpc, this gain in
significance is almost precisely compensated by the extra distance-cut
trial correction. \black

\begin{figure}
\begin{center}
\includegraphics[width=0.8\columnwidth]{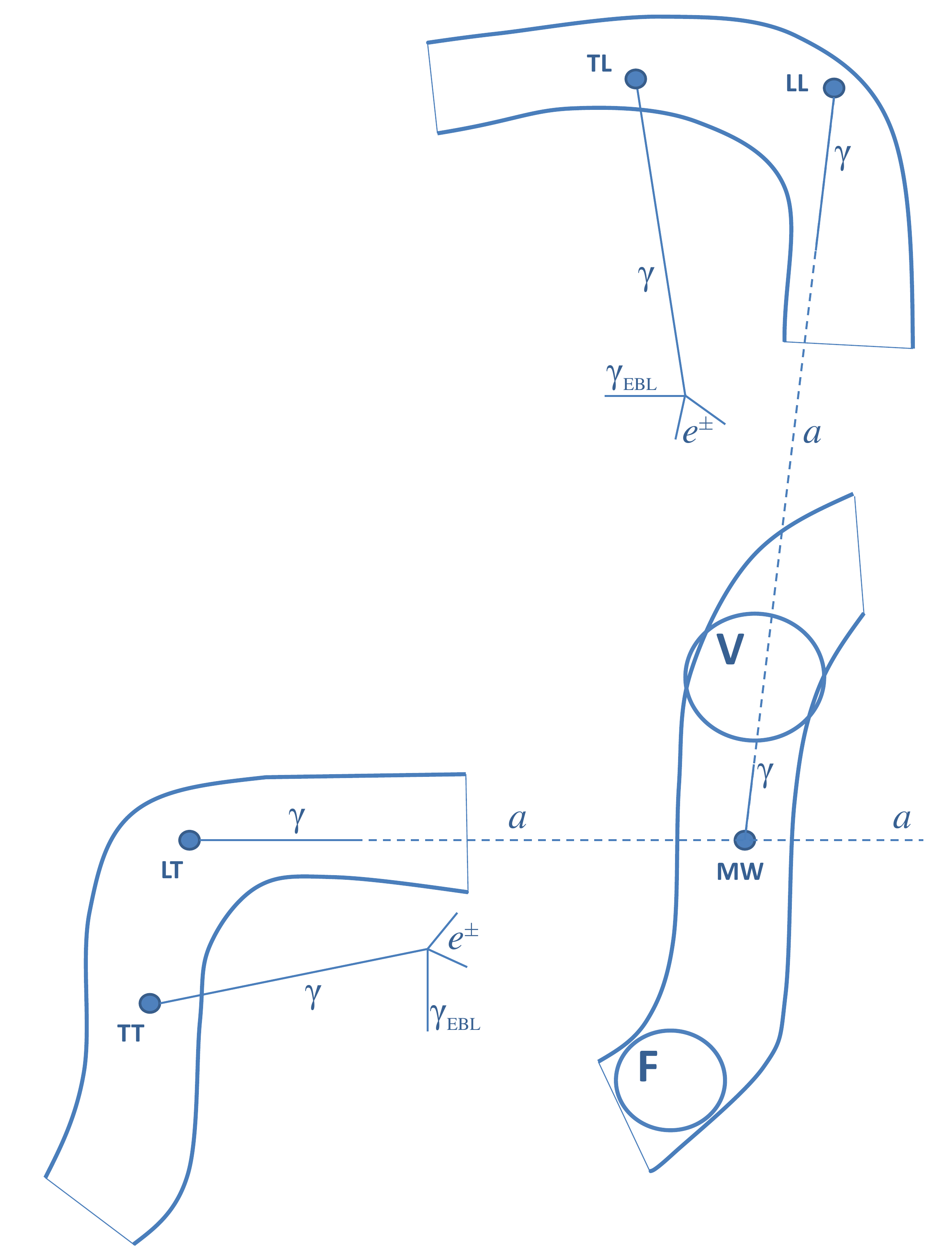}
\caption{
\label{fig:sketch}
A sketch (not to scale) of the explanation of the observed
anisotropy involving ALP-photon mixing in filaments. The Milky Way (MW) is
in the local filament connecting the Virgo (V) and Fornax (F) clusters.
Blazars also sit in their local filaments. For sources with the line of
sight transversal to their filaments (TT and TL), the photons have no time
to convert to ALPs close to the source. If the line of sight goes along
the blazar filament (LT and LL), a part of the high-energy photons
convert to ALPs there, but only for those coming along our filament, LL,
they have enough time to reconvert back to photons close to the Milky Way.
}
\end{center}
\end{figure}

This \black correlation with the local matter distribution \black fits the
predictions of the model of ALP-photon mixing \cite{FRT}: a part of
emitted photons convert to ALPs in the magnetic field of the filament
containing the source and reconvert back to photons in the filament
containing the observer. The maximal-mixing conditions \cite{FRT} depend
on the ALP mass $m$, the ALP-photon coupling $g$, the photon energy $E$,
the magnetic field $B$ and the size of the field-filled region $L$. The
observed anisotropy suggests that the conversion happens in the directions
along the filament and not in the transverse direction, as it is
graphically shown in Figure~\ref{fig:sketch},
which illustrates the interpretation of our main result in terms of the
photon-axion mixing\footnote{The author thanks O.~Troitskaya for her help
with this sketch.},
so 0.5~Mpc$\lesssim
L \lesssim$20~Mpc. The magnetic field in filaments \cite{MF-in-fil} is
poorly known. Computer simulations \cite{Dolag} indicate that $B\sim
10^{-8}$~G in cluster outskirts. It is quite nontrivial to satisfy the
maximal-mixing conditions for very different energies, $E\sim 10^{12}$~eV
(VHE) and $E\sim 10^{19}$~eV (UHE), simultaneously, especially given that
the ALP-photon mixing is suppressed by Quantum-Electrodynamics effects for
large values of the product $EB$, see Ref.~\cite{FRT} for details.
Remarkably, these conditions are indeed satisfied, for both energy bands,
in the local filament for $g\sim \mbox{(a few)} \times
10^{-11}$~GeV$^{-1}$ and $m\sim \mbox{(a few)} \times 10^{-9}$~eV. It is
hardly posiible to estimate ALP parameters more precisely before a firm
magnetic-field model is constructed. This parameter range is allowed (see
e.g.\ Ref.~\cite{ST-mini}) by present experimental and astrophysical
limits\footnote{Previous claims of the exclusion of a part of this
parameter space from the lack of irregularities in gamma-ray spectra of
sources embedded in the magnetic field of galaxy clusters have been
recently shown to suffer from orders-of-magnitude systematic uncertainties
 \cite{irreg}.}. It narrows the range previously invoked for the
explanation of the anomalous transparency of the Universe, see
e.g.~Ref~\cite{ST-PRD}. At the same time, it is within the reach of
experiments of the near future, including solar axion telescopes TASTE
 \cite{TASTE} and IAXO  \cite{IAXO}, as well as the laboratory experiment
ALPS-IIc  \cite{ALPS-IIc}.

Because of small fields
of view of IACTs, no survey of a significant part of the sky is available
at TeV energies.
Of the three lists we used for construction of control samples,
the 3FHL catalog and the list of HiRes stereo events represent complete
samples (that is, follow a known sky coverage). However, the TeVCat sample
is constructed on the base of individual observations while non-detections
are normally not published and the sky coverage cannot be quantitatively
determined. Hence, it is not complete in this sense and may be biased. For
the present analysis, however, this possible bias is conservative. Indeed,
the anomalous blazars are defined in our VHE sample as those whose
deabsorbed spectra harden right at the energy where the deabsorption
correction becomes essential. We assume that the true intrinsic spectra of
blazars are either power-law or concave since no physical reason for
convexity at these energies is known. Suppose now that there is a region
in the sky in which direction the Universe is anomalously transparent at
high energies, that is the standard deabsorption results in too high
fluxes at high energies (hardenings). Then those spectra which are
power-law would be reconstructed with upward breaks; they join our
signal sample $S_{V}$. Intrinsically concave spectra would not
exhibit hardening after deabsorption, but these objects would look
brighter than physically similar sources observed in other directions in
the sky. These stronger high-energy fluxes would make them more probable
to enter the sample which consists of detected objects only.
Interestingly, this is right what we observe: in
Fig.~\ref{fig:mapPBR-dots}, objects from the TeVCat sample without
anomalies, represented by black dots, also have a tendency to concentrate
towards directions with larger values of $f(l,b)$. In this context it is
interesting to note that all three gamma-ray bursts detected at VHE and
listed in TeVCat are also seen in the direction of the local filament, see
Fig.~\ref{fig:mapBLandPBRandLSS} (we did not study whether they exhibit
any kind of anomalies in their spectra). While the tendency is much weaker
for the objects without hardenings than the effect we find for
$S_{V}$, it explains the conservative character of the TeVCat
selection bias: the control sample contains many ``signal'' objects. We
believe that this is the reason for the p-value for $S_{V}$ to
be considerably larger than that for $S_{U}$.

Future tests of the observed effect at VHE are expected to overcome
this conservative bias.
This may be cured when a more complete sample of VHE
blazars, incuding non-detections, will be available, for instance, from
the Cherenkov Telescope Array~\cite{CTA}. Another option is to develop a
more sensitive method for the Fermi-LAT data. This instrument has a small
effective area, as compared to IACTs, and presently only five blazars are
significantly detected by LAT at the energies where the absorption is
important~\cite{gamma2}. Other statistical methods than used here should
be applied in order to benefit from the Fermi-LAT full sky coverage. In
future, the sensitivity in the Fermi-LAT band might be improved with
low-threshold high-altitude IACTs  \cite{5x5}, e.g.\ ALEGRO
\cite{ALEGRO}.

Similarly to the VHE case, the list of BL~Lacs used in
Refs.~\cite{HiRes271,BLL-HiRes} is not complete: they were selected from
serendipituous studies. While it is hard to account for this
incompleteness, it can hardly have any relation to the local filament.

Further tests of the BL Lac/UHE-shower correlations are necessary before a
study of the anisotropy of the effect can be performed with new data. Poor
angular resolution of modern UHE cosmic-ray experiments\footnote{The
resolution of both the Pierre Auger Observatory (Auger) and the Telescope
Array (TA) experiments are roughly twice worse than that of HiRes stereo.}
makes it hard to test the correlation found in the HiRes data, though this
may be partly compensated by \black large \black  statistics.
\black Corresponding estimates are given in Ref.~\cite{BLfuture}, assuming
that the particles which caused air showers correlated with BL Lacs are
similar to the bulk of the cosmic-ray particles with respect to their
detection. Though the actual angular resolution of modern experiments
differs from the one assumed at that time, notably for the hybrid mode,
these numbers can be easily corrected.
Unfortunately, the exceptional character of air showers
induced by primary photons
(or by any other hypothetical unusual particles
-- standard cosmic rays cannot correlate with distant BL Lacs)
makes this
approximation only partially relevant, while moving beyond the
approximation requires internal information about experiments.

Indeed, the correlation was found in the data of a fluorescent detector
(FD)  working in the stereo mode. In principle, it should be tested with
similar data. But Auger does not have the FD stereo reconstruction at all,
while TA has quite low angular resolution in stereo: new FD telescopes are
at larger distances from each other, compared to HiRes. When surface
detectors (SD) are involved, one should take into account their different
responses to unusual air showers \red (which are implied by the HiRes
observation: neutral particles are required)\black. When a correlation
study is performed, potentially unusual events are compared to the bulk of
cosmic rays. If the energy assigned to a primary photon by means of the
bulk reconstruction is underestimated by a factor of $x$, then the
background for the correlation studies would be $x^{\alpha}$ times larger,
where $\alpha\sim 2$ is the index of the power-law integral spectrum. A
correct account of this effect, which is in addition direction-dependent,
requires the use of detailed information about experiments which is not
publicly available.\black

\red The only attempt to test the HiRes BL Lac correlations with
independent data was reported by Auger at a conference in
2007 \cite{Auger-BL}. The Auger SD data were examined and no correlations
with southern-sky BL~Lacs were found. This analysis, in addition to the
low statistics used, faces the SD problem discussed above. For instance,
for photon-induced showers, \black the low response of water tanks
leads to underestimation of the energies of primary photons, $x \gtrsim 3$
\black \cite{Billoir,Nuhuil}; see Ref.~\cite{FRT} for a discussion.
Large-statistics tests of the BL Lac correlations with the TA data are a
necessary prerequisite for further studies of the effect reported here at
UHE.

An important implication of the ALP explanation of the UHE correlations
is that the correlated air showers are initiated by primary photons.
Ref.~\cite{BLfuture} determines the fraction of the correlated events
in a HiRes-like data set as 1.5\% to 3.5\% (at 95\% CL).
\black This translates into the flux
$F_{\rm BL}\gtrsim 5 \times 10^{-3}$~km$^{-2}$yr$^{-1}$sr$^{-1}$
for the HiRes field of view. \red We note that these values of
$F_{\rm BL}$ are of the same order as current \black published 95\% CL
limits on the \red isotropic \black gamma-ray flux from Auger,
$F_{\gamma}<7\times 10^{-3}$~km$^{-2}$yr$^{-1}$sr$^{-1}$
\cite{AugerGammaFlux}, and TA, $F_{\gamma}<3.6\times
10^{-3}$~km$^{-2}$yr$^{-1}$sr$^{-1}$ for $E>10^{19}$~eV
\cite{TAGammaLimit}, \red and, taken at a face value, are in tension with
the Auger limit of $F_{\gamma}\lesssim 2.2\times
10^{-3}$~km$^{-2}$yr$^{-1}$sr$^{-1}$ reported at a conference in 2019
\cite{Auger-ICRC2019-gamma}. However, $F_{\rm BL}$ is a different quantity
than $F_{\gamma}$ which assumes isotropic diffuse flux, so \black
a dedicated study is
required to test the filament hypothesis. \black While independent from
the anisotropy result presented in this paper, further searches for UHE
photons would be important for its interpretation.

\appendix
\section{Construction of the weighted density of galaxies}
\label{app:magic-numbers}
\begin{figure}[b]
\begin{center}
\includegraphics[width=0.9\columnwidth]{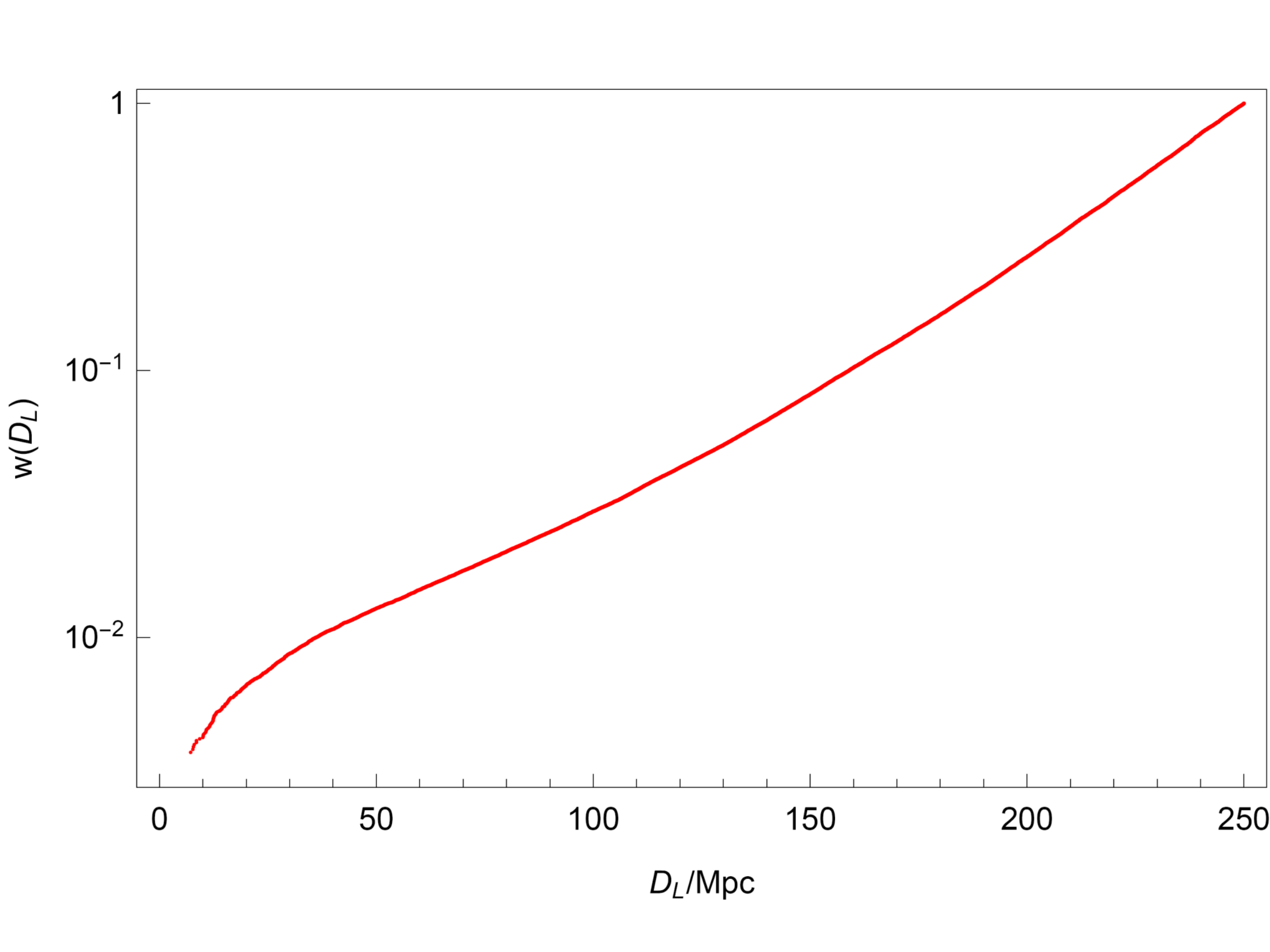}
\caption{
\label{fig:weights}
The weight functon $w(D_{L})$ accounting for the incompleteness of the
flux-limited 2MRS (2019) catalog. }
\end{center}
\end{figure}

For the construction of the weighted density of galaxies in the local
Universe, we follow the approaches developed and applied previously in the
UHECR context, see e.g.\ Refs.~\cite{UHE-LSS1,UHE-LSS2,UHE-LSS3}. We
start from the flux-limited ($K_{S}<11.25$) complete catalog of galaxies
with coordinates and radial velocities, the 2MASS Redshift Survey (2MRS)
in its 2019 release, Refs.~\cite{2MASS1,2MASS2}. Like in
Ref.~\cite{UHE-LSS3}, we select galaxies with
luminosity distances\footnote{The catalog gives radial velocities which we
convert to distances assuming the flat $\Lambda$CDM cosmological model
with $\Omega_M=0.308$, $\Omega_{\Lambda}=0.692$, $H_{0}=67.8$~km/s/Mpc.}
$D_{L}$ between 5~Mpc and 250~Mpc and attribute a weight to each galaxy.
This weight is a product of $1/D_{L}^{2}$ and an additional factor which
accounts for progressive incompleteness of the flux-limited catalog at
large distances. This latter factor is calculated with the help of the
``sliding-box'' method described in detail in Ref.~\cite{Hylke}, adopting
in turn Ref.~\cite{sliding-old}. Figure~\ref{fig:weights}
presents this factor as a function of $D_{L}$.

In this way, we obtain a list of galaxies with their Galactic coordinates
$(l_{i},b_{i})$ in the sky and their weights,
$w_{i}=w(D_{L,i})/D_{L,i}^{2}$, $i=1, \dots, N_{\rm tot}$, where $N_{\rm
tot}=41706$ is the total number of galaxies with 5~Mpc$\le D_{L} \le
250$~Mpc and $K_{S} \le 11.25$ in the 2MRS catalog. The projected weighted
density used in our analysis is a function of coordinates $(l,b)$
determined as a sum of these weights over all galaxies in a
(smoothed) given direction. We consider two options of smoothing:
the disk smoothing, equivalent to simple summing of $w_{i}$ within a cone
of the opening angle $\theta_{c}$ centered on $(l,b)$, so that the
additional weight is
$$
\bar w_d (\theta) =
\left\{
\begin{array}{cl}
1,& \theta \le \theta_c;\\
0,& \theta > \theta_c,
\end{array}
\right.
$$
where $\theta$ is the angle between the directions $(l,b)$ and
$(l_{i},b_{i})$; and the Gaussian-like smoothing with the additional
weight function
$$
\bar w_G (\theta) =
\left\{
\begin{array}{cl}
\displaystyle
\frac{1}{2\pi (1-\cos\theta_c)}
\exp{\left(-\frac{1-\cos\theta}{1-\cos\theta_c} \right)},& \theta \le
3\theta_c;\\
0,& \theta > 3 \theta_c.
\end{array}
\right.
$$
Finally, we need to account for the zone of avoidance: the 2MRS catalog is
complete for Galactic latitudes $|b| \ge 8^{\circ}$ in the direction to
the Galactic center, $|l| \le 30^{\circ}$, and for $|b| \ge 5^{\circ}$ for
other Galactic longuitudes $l$. We introduce an additional mask-related
weight for every direction $(l,b)$ equal to the inverse integral of $\bar
w_{d}$ or $\bar w_{G}$ over the sky. If this integral is zero, which may
happen for small values of $\theta_{c}$ and a narrow band near the Galactic
plane, the direction is dropped from the analysis. Various values of
$\theta_{c}$ were used for this study, see Appendix~\ref{app:trials}.

\section{Coordinates of anomalous sources}
\label{app:catalogs}
Tables \ref{tab:pbr} and \ref{tab:bll}  give the lists of anomalous
sources used in the study.
\begin{table}[h]
\begin{center}
\begin{tabular}{cccc}
\hline
$l$    &  $b$  &  $z$   & name \\
\hline
\hline
 158.6 &  47.9 & 0.896  & 4C~55.17   \\
 123.7 &  58.8 & 0.847  & PG~1246$+$586\\
 199.4 &  78.4 & 0.725  & Ton~599     \\
  29.5 &  68.2 & 0.605  & PKS~1424$+$240\\
 305.1 &  57.1 & 0.536  & 3C~279     \\
 191.8 & $-$33.2 & 0.287  & 1ES~0414$+$009\\
 166.2 &  32.9 & 0.138  & 1ES~0806$+$524\\
 188.9 &  82.1 & 0.13   & 1ES~1215$+$303\\
  17.7 & $-$52.2 & 0.116  & PKS~2155$-$304\\
 350.4 & $-$32.6 & 0.071  & PKS~2005$-$489\\
\hline
\end{tabular}
\end{center}
\caption{
\label{tab:pbr}
Galactic coordinates (in degrees), redshifts and names of 10 VHE blazars
with anomalous hardenings.
}
\end{table}~%
\begin{table}[h]
\begin{center}
\begin{tabular}{cccc}
\hline
$l$    &  $b$  &  $z$   & name \\
\hline
\hline
 107.4 &  55.8 & 0.690  & RX~J1359.8$+$5911 \\
 170.5 &  30.1 & 0.377  & TXS~0751$+$485 \\
 211.2 &  69.0 & 0.360  & RX~J1117.6$+$2548 \\
 191.1 &  42.5 & 0.354  & Ton~1015   \\
 141.6 & $-$75.1 & 0.234  & RBS~161   \\
 160.2 &  58.2 & 0.140  & RGB~J1053$+$494 \\
  98.0 &  17.7 & 0.047  & 1ES~1959$+$650   \\
 174.2 & $-$41.9 &   -    & RBS~400   \\
 160.0 &  33.4 &   -    & RGB~J0816$+$576 \\
  91.8 &  52.0 &   -    & SBS~1508$+$561   \\
  64.4 &  39.1 &   -    & RGB~J1652$+$403   \\
\hline
\end{tabular}
\end{center}
\caption{
\label{tab:bll}
Galactic coordinates (in degrees), redshifts (when measured) and names of
11 BL Lac type objects correlating with UHE air showers. }
\end{table}

\section{Account of multiple trials}
\label{app:trials}
\begin{figure}
\begin{center}
\includegraphics[width=0.9\columnwidth]{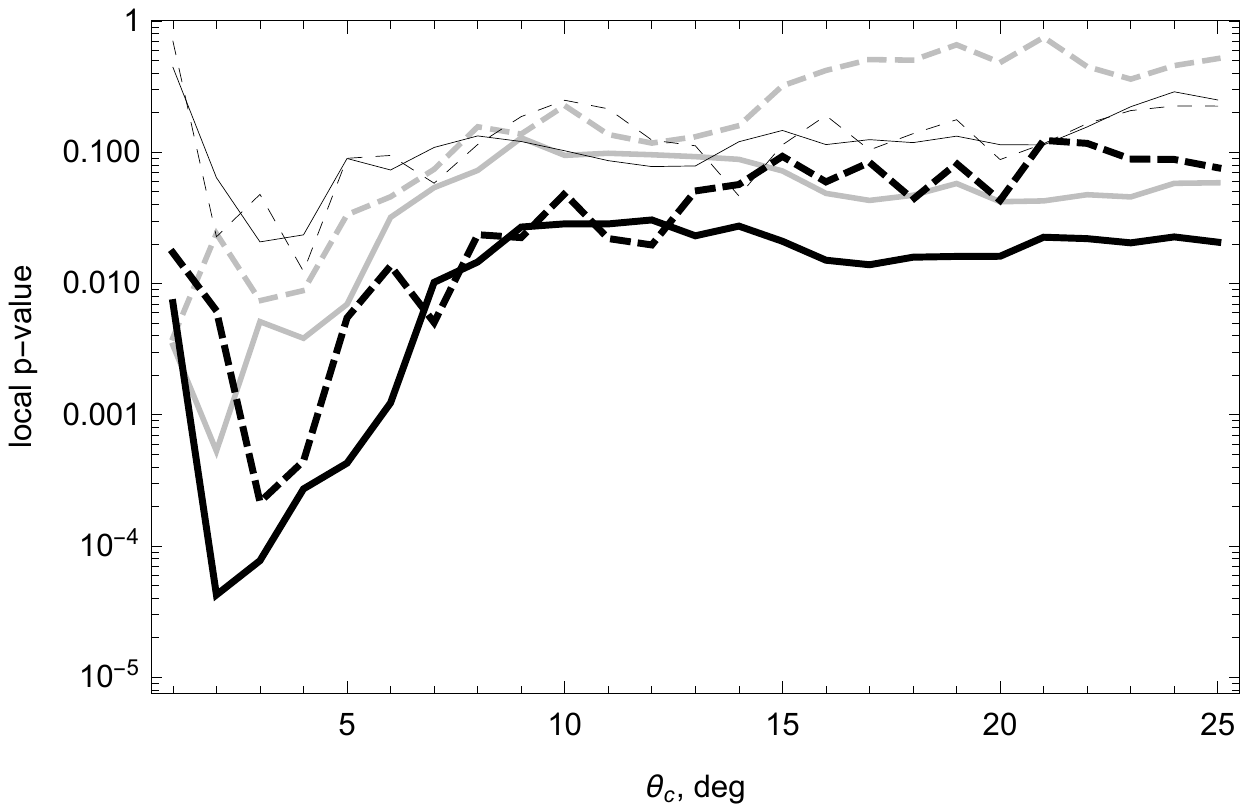}
\caption{
\label{fig:trials}
The local p-value as a function of the free parameter $\theta_{c}$ and
the choice of the disk (dashed) or Gaussian (full) smoothing. Thin black
lines -- VHE hardenings, gray lines -- UHE correlations, thick black lines
-- combined analysis. }
\end{center}
\end{figure}

In the present study, we do not assume any particular quantitative model of
the anomalous transparency. \green While our results support the photon-ALP
conversion on the magnetic field in the local filament, this field is
poorly known. To which extent the number density of galaxies is a tracer
of the filament field is also unknown. Quantitatively, this lack of direct
relation is parametrized by the angular smoothing we introduce in the
calculation of the weighted density. \black We choose to scan over the
smoothing angular scale, $1^{\circ} \le \theta_{c} \le 25^{\circ}$ in
steps of $1^{\circ}$, and to try both ways of smoothing described in
Appendix~A, treating the scan in the unknown parameter as multiple trials
in our statistical study. As it is customary for the account of multiple
trials, we first calculate the local p-value for each of the 50 variants
of smoothing (two functions $\bar w_{d,G}$ and 25 values of $\theta_{c}$),
see Fig.~\ref{fig:trials}.
The minimal p-value, obtained for a certain variant of smoothing, is
called the pre-trial p-value, $p_{1}$.

The next step is necessary to directly estimate how often this or lower
p-value can appear as a fluctuation because of the large number of trials.
Were the trials statistically independent, this step would result in the
multiplication of $p_{1}$ by the number of trials, 50 in our case.
However, this is not the case here: different trials correspond to
different versions of smoothing of one and the same distribution and hence
are strongly statistically interdependent. We use the standard method of
treating the multiple comparisons issue for non-independent data, see
e.g.\ Ref.~\cite{TT-penalty}. We generate a large number, $N=10^{6}$,
Monte-Carlo samples which imitate the sample of anomalous sources, and
repeat the procedure of calculating the local p-values and taking the
minimal of them, $p_{1,k}$, for each sample $k$. The number $M$ of
random samples for which $p_{1,k}\le p_{1}$ determines the global, or
post-trial, p-value, $p\approx M/N$. This $p$ is interpreted as the
probability that the distribution of weighted densities of galaxies in the
directions to anomalous sources differs from that for all sources due to a
random fluctuation for any assumed smoothing function. By definition, this
is the probability that determines the significance of the rejection of
the null hypothesis of isotropy in our analysis.

\begin{acknowledgements}
The author is indebted to
A.~Korochkin, M.~Kuznetsov, M.~Libanov,
A.~Plavin,
M.~Pshirkov,
V.~Rubakov, G.~Rubtsov and P.~Tinyakov
for
illuminating and helpful discussions of various aspects of this study.
\end{acknowledgements}



\end{document}